\documentclass[a4paper,UKenglish,cleveref,autoref,thm-restate]{lipics-v2021}
\hideLIPIcs
\nolinenumbers

\usepackage{xspace}
\usepackage{algorithm}
\usepackage[noend]{algpseudocode}
\usepackage{booktabs}
\usepackage{pgfplots}
\pgfplotsset{compat=newest}

\usepgfplotslibrary{groupplots}
\usetikzlibrary{pgfplots.polar}
\usetikzlibrary{pgfplots.statistics}
\usetikzlibrary{arrows.meta}
\usetikzlibrary{overlay-beamer-styles}

\definecolor{veryLightGrey}{HTML}{F2F2F2}
\definecolor{lightGrey}{HTML}{DDDDDD}
\definecolor{colorThresholdBased}{HTML}{F8BA01}
\definecolor{colorThresholdBasedConsensus}{HTML}{FF7F00}
\definecolor{colorThresholdBasedBuRR}{HTML}{A65628}
\definecolor{colorThresholdBasedOriginal}{HTML}{F781BF}
\definecolor{colorPaCHash}{HTML}{4DAF4A}
\definecolor{colorPhobicCompact}{HTML}{377EB8}
\definecolor{colorPhobicRice}{HTML}{984EA3}
\definecolor{colorRecSplit}{HTML}{E41A1C}
\definecolor{colorChd}{HTML}{444444}

\pgfdeclareplotmark{flippedTriangle}{%
  \pgfpathmoveto{\pgfqpointpolar{-90}{1.2\pgfplotmarksize}}%
  \pgfpathlineto{\pgfqpointpolar{30}{1.2\pgfplotmarksize}}%
  \pgfpathlineto{\pgfqpointpolar{150}{1.2\pgfplotmarksize}}%
  \pgfpathclose%
  \pgfusepath{stroke}
}
\pgfdeclareplotmark{rightTriangle}{%
  \pgfpathmoveto{\pgfqpointpolar{0}{1.2\pgfplotmarksize}}%
  \pgfpathlineto{\pgfqpointpolar{120}{1.2\pgfplotmarksize}}%
  \pgfpathlineto{\pgfqpointpolar{240}{1.2\pgfplotmarksize}}%
  \pgfpathclose%
  \pgfusepath{stroke}%
}
\pgfdeclareplotmark{leftTriangle}{%
  \pgfpathmoveto{\pgfqpointpolar{-180}{1.2\pgfplotmarksize}}%
  \pgfpathlineto{\pgfqpointpolar{-60}{1.2\pgfplotmarksize}}%
  \pgfpathlineto{\pgfqpointpolar{60}{1.2\pgfplotmarksize}}%
  \pgfpathclose%
  \pgfusepath{stroke}%
}

\pgfplotscreateplotcyclelist{myColorList}{%
  colorThresholdBased,mark=triangle\\
  colorPaCHash,mark=pentagon\\%
  colorPhobicRice,mark=square\\%
  colorRecSplit,mark=o\\%
  colorChd,mark=x\\%
  colorThresholdBasedOriginal,mark=asterisk\\%
}

\pgfplotsset{
  mark repeat*/.style={
    scatter,
    scatter src=x,
    scatter/@pre marker code/.code={
      \pgfmathtruncatemacro\usemark{
        or(mod(\coordindex,#1)==0, (\coordindex==(\numcoords-1))
      }
      \ifnum\usemark=0
        \pgfplotsset{mark=none}
      \fi
    },
    scatter/@post marker code/.code={}
  },
  log x ticks with fixed point/.style={
      xticklabel={
        \pgfkeys{/pgf/fpu=true}
        \pgfmathparse{exp(\tick)}%
        \pgfmathprintnumber[fixed relative, precision=3]{\pgfmathresult}
        \pgfkeys{/pgf/fpu=false}
      }
  },
  log y ticks with fixed point/.style={
      yticklabel={
        \pgfkeys{/pgf/fpu=true}
        \pgfmathparse{exp(\tick)}%
        \pgfmathprintnumber[fixed relative, precision=3]{\pgfmathresult}
        \pgfkeys{/pgf/fpu=false}
      }
  },
  log x ticks with percentage/.style={
      xticklabel={
        \pgfkeys{/pgf/fpu=true}
        \pgfmathparse{exp(\tick)}%
        \pgfmathprintnumber[fixed relative, precision=3]{\pgfmathresult}%
        \%%
        \pgfkeys{/pgf/fpu=false}
      }
  },
  every axis/.style={scale only axis},
  major grid style={thin,dotted},
  minor grid style={thin,dotted},
  ymajorgrids,
  yminorgrids,
  every axis/.append style={
    line width=0.9pt,
    tick style={
      line cap=round,
      thin,
      major tick length=4pt,
      minor tick length=2pt,
    },
    mark options={solid},
  },
  legend cell align=left,
  legend style={
    line width=0.7pt,
    /tikz/every even column/.append style={column sep=2mm,black},
    /tikz/every odd column/.append style={black},
    mark options={solid},
    font=\footnotesize,
  },
  title style={yshift=-2pt},
  enlarge x limits=0.04,
  scale only axis,
  /pgf/number format/1000 sep={},
  axis lines*=left,
  xlabel near ticks,
  ylabel near ticks,
  axis lines*=left,
  label style={font=\footnotesize},
  tick label style={font=\footnotesize},
  every axis y label/.append style={yshift=-1pt,inner sep=0,outer sep=0},
  cycle list name=myColorList,
  construction/.style={
      width=4.2cm,
      height=3.5cm,
      xmin=8,
      xmax=2000,
      ymax=28,
      log x ticks with percentage,
  },
  query/.style={
      width=4.2cm,
      height=3.5cm,
      xmin=8,
      xmax=2000,
      ymax=50,
      log x ticks with percentage,
  },
  dominanceMap/.style={
      patch,
      patch type=rectangle,
      shader=flat corner,
  },
}

\title{Engineering Minimal $k$-Perfect Hash Functions} %

\newcommand{\psfrage}[1]{{\color{blue}{\sf[PS: #1]}}} %
\newcommand{\hpfrage}[1]{{\color{violet}\sf[HP: #1]}} %
\newcommand{\skfrage}[1]{{\color{teal}\sf[SK: #1]}} %
\newcommand{\shfrage}[1]{{\color{brown}\sf[SH: #1]}} %
\newcommand{\swfrage}[1]{{\color{red}\sf[SW: #1]}} %
\renewcommand{\psfrage}[1]{} \renewcommand{\hpfrage}[1]{} \renewcommand{\skfrage}[1]{}\renewcommand{\shfrage}[1]{}\renewcommand{\swfrage}[1]{}

\newcommand{\myparagraph}[1]{\subparagraph*{#1}}
\def\consensus{\texorpdfstring{C\scalebox{0.8}{ONSENSUS}}{CONSENSUS}\xspace}
\newcommand{\etal}{\mbox{et al.}\xspace} %
\let\oldcite\cite
\renewcommand\cite{\unskip~\oldcite}

\author{Stefan Hermann}{Karlsruhe Institute of Technology}{hermann@kit.edu}{https://orcid.org/0000-0001-9183-2926}{}
\author{Sebastian Kirmayer}{Karlsruhe Institute of Technology}{}{}{}
\author{Hans-Peter Lehmann}{Karlsruhe Institute of Technology}{hans-peter.lehmann@kit.edu}{https://orcid.org/0000-0002-0474-1805}{}
\author{Peter Sanders}{Karlsruhe Institute of Technology}{sanders@kit.edu}{0000-0003-3330-9349}{}
\author{Stefan Walzer}{Karlsruhe Institute of Technology}{stefan.walzer@kit.edu}{https://orcid.org/0000-0002-6477-0106}{}

\authorrunning{Hermann, Kirmayer, Lehmann, Sanders, Walzer}
\Copyright{Stefan Hermann, Sebastian Kirmayer, Hans-Peter Lehmann, Peter Sanders, Stefan Walzer}

\EventEditors{Anne Benoit, Haim Kaplan, Sebastian Wild, and Grzegorz Herman}
\EventNoEds{4}
\EventLongTitle{33rd Annual European Symposium on Algorithms (ESA 2025)}
\EventShortTitle{ESA 2025}
\EventAcronym{ESA}
\EventYear{2025}
\EventDate{September 15--17, 2025}
\EventLocation{Warsaw, Poland}
\EventLogo{}
\SeriesVolume{351}
\ArticleNo{98}


\supplementdetails{Software}{https://github.com/stefanfred/engineering-k-perfect-hashing} %

\funding{
    \flag[2cm]{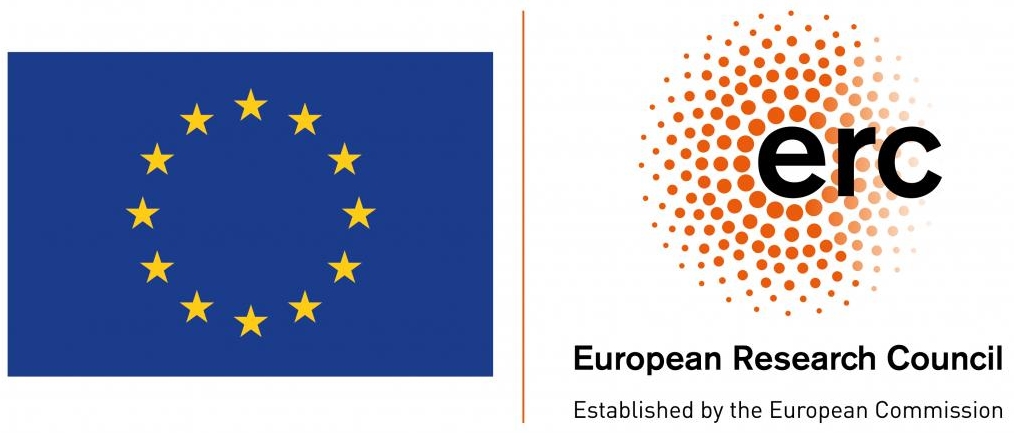}
    This project has received funding from the European Research Council (ERC) under the European Union’s Horizon 2020 research and innovation programme (grant agreement No. 882500).
    This work was also supported by funding from the pilot program Core Informatics at KIT (KiKIT) of the Helmholtz Association (HGF).
}
\acknowledgements{
    This paper is based on and has text overlaps with the bachelor's thesis of Sebastian Kirmayer \cite{kirmayer2024engineering}.
}

\newcommand{\fvref}[1]{in \cref{#1}}

\begin{document}

\maketitle

\begin{abstract}
    Given a set $S$ of $n$ keys, a $k$-perfect hash function ($k$PHF) is a data structure that maps the keys to the first $m$ integers, where each output integer can be hit by at most $k$ input keys.
    When $m=\lceil n/k \rceil$, the resulting function is called a \emph{minimal} $k$-perfect hash function (M$k$PHF).
    Applications of $k$PHFs can be found in external memory data structures or to create efficient 1-perfect hash functions, which in turn have a wide range of applications from databases to bioinformatics.

    Several papers from the 1980s look at external memory data structures with small internal memory indexes.
    However, actual $k$-perfect hash functions are surprisingly rare, and the area has not seen a lot of research recently.
    At the same time, recent research in 1-perfect hashing shows that there is a lack of efficient $k$PHFs.
    In this paper, we revive the area of $k$-perfect hashing, presenting four new constructions.
    Our implementations simultaneously dominate older approaches in space consumption, construction time, and query time.
    We see this paper as a possible starting point of an active line of research, similar to the area of 1-perfect hashing.
\end{abstract}

\newpage
\section{Introduction}
Given a set $S$ of $n$ keys, a $k$-perfect hash function ($k$PHF) is a data structure that maps the keys to the first $m$ non-negative integers $[m]$, called \emph{bins} in the following.
Each output bin can be hit by at most $k$ input keys.
The output for any key not in $S$ is undefined, so a $k$-perfect hash function does not need to store the input set.
For $k=1$, there is a wide range of constructions covering different trade-offs between space consumption, query time, and construction time.
The area is an active field of research with several publications each year  \cite{lehmann2024fast, lehmann2025modern,lehmann2023bipartite,lehmann2025consensus,esposito2020recsplit,hermann2024phobic,grootkoerkamp2025ptrhash,beling2023fingerprinting,lehmann2023sichash,hermann2025morphishash,pibiri2021pthash}.
However, there has been little prior work on $k$-perfect hashing.
While several papers from the 1980s look at internal memory indexes for external hash tables, these are not full $k$-perfect hash functions.
We refer to \cref{s:related} for details.
In this paper, we present four new $k$-perfect hash function constructions.
Each covers a different trade-off between the three main performance metrics -- space consumption, construction time, and query time.

\myparagraph{Minimality.}
When $m=\lceil n/k \rceil$, we obtain a \emph{minimal} $k$-perfect hash function (M$k$PHF) \cite{larson1985external}.
When using the M$k$PHF to partition 1-PHFs \cite{lehmann2023bipartite,lehmann2025consensus}, it might be beneficial to look at a stricter definition where all output bins contain exactly $k$ keys, except possibly the last.
The approaches presented in this paper fulfill the stricter definition.
An alternative definition of non-minimal $k$PHFs could be that we could fill each output bin with either exactly $k$ keys or $0$ keys.
For $k=1$, this results in an ordinary non-minimal PHF and achieves a significant reduction in space consumption.
However, for $k>1$, the resulting space consumption is the same as the one of M$k$PHFs \cite{mairson1992effect}.

\myparagraph{Applications.}
$k$-perfect hashing has a wide range of applications.
In external memory hash tables, $k$PHFs can return the page on which a key is stored.
External memory is usually accessed with page granularity anyway, so having to scan the page for the key only causes negligible overhead.
Compared to a 1-PHF, however, this can significantly reduce the space needed.
For internal memory hash tables it can be beneficial to store more than one key in each table cell \cite{fotakis2005space}.
Using an M$k$PHF, we can completely fill such a hash table.
A $k$-perfect hash function can also be used to make the construction of 1-perfect hash functions more uniform by dividing the input to same-size partitions \cite{lehmann2023bipartite,lehmann2025consensus}.

\myparagraph{Our Contributions.}
In this paper, we present four new or significantly improved $k$-perfect hash function constructions.
Threshold-based bumping was previously only briefly described as an ad-hoc solution to aid 1-perfect hashing \cite{lehmann2023bipartite,lehmann2025consensus}.
In \cref{s:thresholdBased}, we extend it through optimal thresholds and \consensus-coded \cite{lehmann2025consensus} thresholds.
Perfect hashing through bucket placement \cite{pibiri2021pthash,hermann2024phobic,belazzougui2009hash} generalizes to $k$-perfect hashing in an obvious way.
In \cref{s:bucketPlacement}, we introduce an optimal bucket assignment function.
RecSplit \cite{esposito2020recsplit} is a known 1-PHF, which we extend to a $k$-PHF in \cref{s:recusiveSplitting}.
Finally, we introduce a variant of the external memory hash table PaCHash \cite{kurpicz2023pachash} that can be used as an M$k$PHF in \cref{s:kperfectPaCHash}.
To the best of our knowledge, we then give the first experimental evaluation of different $k$-perfect hash functions in \cref{s:eval}.

\section{Preliminaries}
In this section we show the space lower bound for minimal $k$-perfect hash functions.
We then explain preliminary data structures that we use throughout this paper.

\myparagraph{Space Lower Bounds.}\label{s:spacebounds}
The space lower bound for representing a minimal $k$-perfect hash function is rather easy to prove \cite{mairson1983program,belazzougui2009hash,kurpicz2023pachash}.
Let the $n$ input keys be a subset of a universe of size $u$. 
There are ${u \choose n}$ possible input sets.
Each behavior of an M$k$PHF can cover at most ${u/(n/k) \choose k}^{n/k}$ input sets.
Therefore, the number of bits we need is at least
\begin{align*}
\hspace{-2mm}
\log_2\left(\frac{{u \choose n}}{{u/(n/k) \choose k}^{n/k}}\right)
        \stackrel{n^2 \in o(u)}{\approx}  n \cdot \left(\log_2(e) - \log_2(k^k/k!)/k\right)
        = n\cdot\left(\frac{\log_2(2 \pi k)}{2k} + \Theta\left(\frac{1}{k^2}\right)\right).
\end{align*} 
In \cref{tab:bounds}, we give a range of example values.
The space lower bounds become more complex for non-minimal $k$-perfect hashing, and no tight bounds are known yet \cite{belazzougui2009hash}.
\begin{table}[t]
  \caption{%
      Space lower bound for different values of $k$.
  }
  \label{tab:bounds}
  \centering
  \begin{tabular}[t]{lrrrrrr}
    \toprule
      $k$          & 1     & 2     & 4     & 10    & 100   & 1000  \\
      Bits per key & 1.443 & 0.943 & 0.589 & 0.300 & 0.046 & 0.006 \\
    \bottomrule
  \end{tabular}
\end{table}

\myparagraph{Retrieval Data Structures.}\label{s:retrieval}
A retrieval data structure stores a static function $f : S \rightarrow \{0, 1\}^r$ with $r$-bit output values.
It can return arbitrary results for any key not in $S$.
This makes it possible to represent it using just $rn$ bits.
Bumped Ribbon Retrieval (BuRR) \cite{dillinger2022burr} is based on solving a system of linear equations and has a space consumption of $1.01rn$ bits in practice.

\myparagraph{Elias-Fano Coding.}\label{s:eliasFano}
Let $(a_i)_{i \in [n]}$ be a monotonically increasing sequence of integers.
To store this sequence with Elias-Fano coding \cite{elias74efficient,fano71number}, we split each integer into two parts.
We store the lower $L = \lceil \log_2(a_{n-1}/n) \rceil$ bits directly in a packed array.
Then we store the remaining upper bits $(h_i)_{i \in [n]}$ as a bit vector where the bits at indices $h_i + i$ are 1, and all other bits are 0.
On this bit vector, we have $h_i = \textrm{select}_1(i) - i$.
Together with a lookup in the packed array, we get constant access time.
In total, storing the sequence needs $n \cdot (2 + \lceil\log_2(a_{n-1}/n)\rceil) + o(n)$ bits of space \cite{elias74efficient,fano71number}.

\myparagraph{Golomb-Rice Coding.}\label{s:golombRice}
For a parameter $L$, Golomb-Rice coding \cite{golomb1966run,rice1979some} divides an integer $x$ into the two parts $h = \lfloor x/2^L \rfloor$ and $l = x \textrm{ mod } 2^L$.
It encodes $l$ using $L$ bits in binary coding, and $h$ in unary coding.
Golomb-Rice coding is optimal for geometrically distributed $x$ with $p=2^{-L}$.
We can store a sequence of integers with constant time access by using a common array for the lower bits, concatenating the upper bits, and using a select data structure.

\section{Related Work}\label{s:related}
We now explain $k$-perfect hash functions and similar external memory hashing data structures from the literature.
Note that Alon \etal \cite{alon1995color} and Berman \etal \cite{berman1986collections} use the variable $k$ for the range of the output values in 1-perfect hashing, which is now usually called $m$.
In an online setting when building hash tables, Frei and Wehner \cite{frei2023bounds} call a family of hash functions \emph{$k$-ideal} if there is, for any input set of size $n$, some hash function that is $k$-perfect.

\subsection{$k$-Perfect Hash Functions}

\myparagraph{Brute-Force Search.}
Similar to what can be done for 1-perfect hashing \cite{melhorn1984data}, we can use brute-force search to construct $k$-perfect hash functions.
We use a hash function that can be parameterized by a seed value, changing its behavior completely.
We greedily start trying seed values until we come across one seed where the hash function is $k$-perfect on our given input set.
Then storing the index of that seed in binary coding is enough to represent the $k$-perfect hash function.
Unfortunately, like for 1-perfect hashing, we need an exponential number of hash function evaluations, making the approach unfeasible for large input sets.
Mairson \cite{mairson1983program} shows that the probability of a random function being an M$k$PHF can be described by the multinomial coefficient
$(n/k)^{-n}{n \choose \underbrace{k,k,\ldots,k}_\textrm{$n/k$ times}}$.

\myparagraph{Bucket Placement.}\label{s:relatedBucketPlacement}
Perfect hashing through bucket placement \cite{pibiri2021pthash,hermann2024phobic,belazzougui2009hash} is a class of 1-perfect hash functions.
The idea is to hash the input keys to small buckets of expected size $\lambda$, where $\lambda$ is a tuning parameter usually in the range 2--6 keys.
For each bucket, we then greedily search for a hash function seed such that the keys in that bucket do not collide with previous keys.
After placing all buckets, we compress the list of seeds while retaining fast access.
The first buckets are easier to place because all bins are empty.
It is therefore helpful to insert the buckets in decreasing order of size.
In CHD \cite{belazzougui2009hash}, the keys are hashed uniformly to the buckets.
Despite sorting, the last buckets are still much harder to place than the first.
Therefore, PHOBIC \cite{hermann2024phobic} introduces an optimal \emph{bucket assignment function} $\beta: [0,1] \rightarrow [0,1]$.
A key with a normalized hash value $x \in [0,1]$ is then assigned to bucket $\lceil\beta(x)n/\lambda\rceil$.
The bucket assignment function makes the first buckets much larger to allow even smaller buckets towards the end, resulting in roughly the same success probability of a seed in all buckets.

Bucket placement can be used as a $k$-perfect hash function by allowing up to $k$ collisions for each output bin.
The implementation \cite{cmph} of CHD supports only non-minimal $k$-perfect hashing.
In \cref{s:bucketPlacement}, we extend the idea through an optimized bucket assignment function.

\myparagraph{Predecessor Dictionary.}\label{s:predecessorDictionary}
Pagh \cite{pagh2003basic} suggests hashing all input keys with a hash function of range $[n^3]$ and sorting them by hash value.
Now map the keys linearly to the output bins and store the smallest hash value in each bin.
For a given hash value $x$, a \emph{predecessor query} can give the largest value $y$ in the list such that $y \leq x$.
The index of that value is the $k$-perfect output bin.
We implement a variant in \cref{s:kperfectPaCHash}.

\myparagraph{Threshold-Based Bumping.}\label{s:thresholdBasedRelWork}
Threshold-based bumping \cite{lehmann2023bipartite} is an ad-hoc solution to $k$-perfect hashing introduced for simplifying a 1-perfect hash function.
The idea is to first hash all keys to $\lceil 0.9 n/k \rceil$ bins.
The constant of $0.9$ is selected arbitrarily in the paper, ensuring that there are few bins that receive fewer than $k$ keys.
An additional hash function uniformly assigns a fingerprint value $\in (0, 1]$ to each key.
For each bin, threshold-based bumping then stores a threshold such that at most $k$ keys in the bin have fingerprints below or equal to the threshold.
It \emph{bumps} keys with fingerprints larger than the threshold to a second layer of the same data structure.
To reduce the space consumption and to enable fast access, we restrict ourselves to a small list of possible thresholds and only store the index into the list.
Threshold-based bumping has similarities with perfect hashing through fingerprinting \cite{muller2014retrieval} (which essentially uses 1-bit thresholds) and separator hashing \cite{gonnet1988external,larson1984file} (see \cref{s:separator}).
The idea of bumping is also used in BuRR \cite{dillinger2022burr}, which solves a system of linear equations and bumps conflicting rows.

To make the resulting function \emph{minimal} $k$-perfect, keys bumped in the second layer have to be placed into the empty slots left in the bins.
Threshold-based bumping does that by storing an Elias-Fano coded list of bins that still have slots left.
Using a minimal 1-PHF on the bumped keys, it can map each key to one of the empty slots.
Refer to \cref{fig:thresholdBased} for an illustration.
Queries hash the key to a bin.
Most keys can immediately return the bin index because their fingerprint is below the threshold.
Some have to look at the next layer and few have to query the fallback 1-PHF.
In \cref{s:thresholdBased}, we extend the approach through an optimal selection of threshold values, as well as more efficient overloading to minimize the number of bins with less than $k$ keys.

\subsection{External Memory Hash Tables}
In external memory hash tables, a goal is to find objects while only accessing a small number of external memory pages.
For this, different internal memory index data structures have been proposed.
These index data structures are very similar to $k$-perfect hash functions.
The main difference is that they either need to inspect more than one external memory page or that they do not fully utilize the external memory, meaning that they are not \emph{minimal} $k$-perfect.
Still, these data structures use similar ideas as M$k$PHFs or can even be adapted to M$k$PHFs.

\myparagraph{Separator Hashing.}\label{s:separator}
Separator hashing \cite{gonnet1988external,larson1984file,larson1988linear} is a static, external memory hash table that guarantees a single disk access for each query.
Each key has a \emph{probe sequence} of output bins and a \emph{fingerprint sequence}, both determined through a sequence of hash functions.
For each external memory page, it stores a threshold value for the fingerprint, also called \emph{separator}.
When querying a key, we look at the bins along its probe sequence.
If the key's fingerprint for that bin is smaller than the separator, we found the corresponding page.
Otherwise, we continue scanning.
A problem with the technique, especially for high load factors, is that bumped keys map back into the same range of bins.
This causes a series of dependent memory accesses during queries, and complicates the construction.

\myparagraph{External Robin-Hood Hashing.}
Instead of displacing keys based on their fingerprint, \emph{external robin-hood hashing} \cite{celia1988external} displaces the keys that are at the smallest index of their probe sequence.
The internal memory then stores, for each page, the smallest distance of a contained key to its home address.
Queries iterate over the distance array starting from their home address until the distance is at least what is stored in the array.
If multiple keys have the same home address, more than one external memory page has to be searched.

\myparagraph{Extendible Hashing.}
The idea of \emph{extendible hashing} \cite{fagin1979extendible} is to store an internal memory table holding pointers.
Each row points to an external memory page holding the keys hashed to that table row.
If the keys hashed to several (adjacent) table rows fit onto an external memory page together, extendible hashing stores the same pointers multiple times.
The approach achieves a load factor of about 69\% \cite{pagh2003basic}.

\myparagraph{PaCHash.}\label{s:pachash}
PaCHash \cite{kurpicz2023pachash} is designed as a static external memory hash table with variable-length keys.
However, we only explain the case that each external memory page can hold $k$ fixed-length keys.
Then PaCHash uses a hash function of range $[a \cdot n/k]$, where $a$ is a tuning parameter.
This divides the keys into different buckets.%
\footnote{For consistency within this paper, we interchange the terms \emph{bin} and \emph{bucket} of the PaCHash paper.}
For each external memory page, we store the index of the first bucket overlapping it.
Using Elias-Fano coding (see \cref{s:eliasFano}), this takes $n/k(2+\log(a))$ bits.
To locate a key, we perform a predecessor query on that sequence to determine which external memory page should be loaded.
If a bucket overlaps the page boundaries, the index cannot determine whether the key is on the page or on the previous page.
In that case, which happens for roughly $1/a$ of the queries, PaCHash loads both adjacent pages.
In \cref{s:kperfectPaCHash}, we explain how we can adapt PaCHash to an M$k$PHF.

\section{$k$-Perfect Hashing Through Threshold-Based Bumping}\label{s:thresholdBased}
\begin{figure}[t]
    \centering
    \includegraphics[scale=0.9]{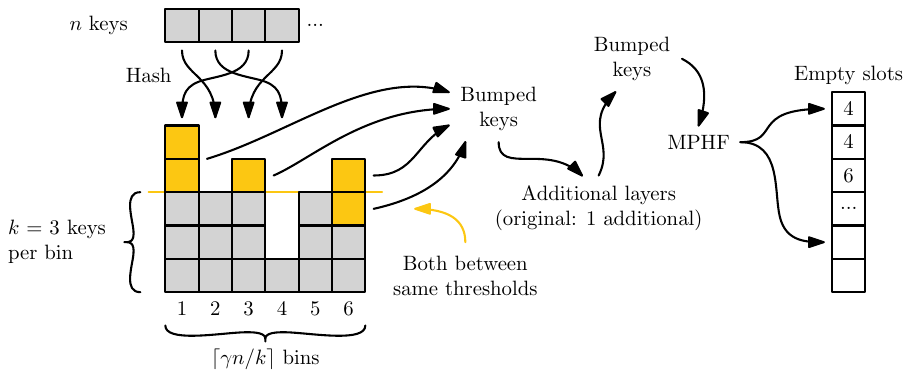}
    \caption{%
        Threshold-based $k$-perfect hashing.
        Figure adapted from \cite{lehmann2023bipartite}.
    }
    \label{fig:thresholdBased}
\end{figure}

Threshold-based bumping is introduced in ShockHash-Flat \cite{lehmann2023bipartite} as an ad-hoc implementation to aid 1-perfect hashing.
In this section, we enhance it significantly.
Remember from \cref{s:thresholdBasedRelWork} that threshold-based bumping hashes each key directly to a bin.
Additionally, we hash each key to determine a uniform fingerprint value $\in (0, 1]$.
Each bin stores a threshold such that at most $k$ keys with fingerprints smaller than or equal to the threshold remain.
The other keys are \emph{bumped} to another layer of the same data structure.
\Cref{fig:thresholdBased} gives an illustration.

\begin{lstlisting}[label=alg:constructionQueryThreshold,float=t,
	caption={Query algorithm of $k$-perfect hashing trough threshold-based bumping. Hash function $h_X$ uniformly maps into the set $X$.}]
Function query$(\textit{key} \in S)$
  $\textit{numBins}$ := $\lceil (n / k) / \gamma \rceil$
  $\textit{bin}$ := $h_{[\textit{numBins}]}(\textit{key})$
  $thresholdIndex$ := &\textit{thresholds}[\textit{bin}]&
  $\textit{fingerprint}$ := $h_{[0,1]}(\textit{key})$
  if $\textit{fingerprint} \leq T_{thresholdIndex}$
    return $\textit{bin}$
  else // bumped to next layer
    return $\textit{numBins} + \text{query next layer}$ &
\end{lstlisting}

\subsection{Overloading}
To make it more likely that each bin receives at least $k$ keys, we \emph{overload} the bins to contain more than $k$ keys in expectation.
While the ad-hoc implementation \cite{lehmann2023bipartite} already overloads the first layer, we now take this to its logical conclusion and overload all layers.
More concretely, we start with $\lceil \frac{n}{k\gamma} \rceil$ bins where $\gamma>1$ is the overloading factor.
We determine the thresholds to store, which leaves us with $n'$ bumped keys.
From the remaining output bins, we then use $\lceil \frac{n'}{k\gamma} \rceil$ bins for the next layer.
We continue this as long as the total number of bins used is still smaller than $\lceil \frac{n}{k} \rceil$.
Only for the remaining keys, we have to use 1-perfect hashing to puzzle them back into the under-filled bins.
Even with overloading, we maintain expected constant query times and expected linear construction times.
Preliminary experiments with $k=1000$ show that overloading all layers reduces the number of keys bumped in the last layer from 0.2\% (with the original two layers) to 0.02\% (with 5 layers).
Because the fallback 1-PHF needs non-negligible space, this reduces the overall space consumption.
In our implementation, we use the 1-PHF FiPS \cite{lehmann2024fast} due to its small constant space overhead when using it for small sets.

\subsection{Optimizing the Thresholds}\label{ss:optthresh}
For fast access we use a fixed number of bits to represent the threshold of each bin, which gives us $t$ possible threshold values.
In the following we consider the value of those $t$ thresholds.
For each bin we then only store the index of the threshold.
We model the fingerprint of a key as a uniformly distributed hash value $x \in (0,1]$.
If $x$ is greater than the threshold of the key's bin, the key is bumped.
It is less likely that we need threshold values close to $0$ because the expected number of keys in each bin is $k\gamma$.
The ad-hoc implementation \cite{lehmann2023bipartite} therefore heuristically uses uniformly spaced thresholds in $[\frac{2}{3}, 1]$.
In this section, we derive an optimal selection of thresholds.

Our goal is to find the $t$ thresholds $T_1 < \dots < T_t \in [0,1]$ such that they minimize the expected number of empty slots.
As a fallback option, we use $T_1 = 0$, bumping all keys.
We also use $T_t=1$.
A value of $T_t<1$ would mean that keys in the range $(T_t, 1]$ always get bumped, which would counteract our increased load factor.
Therefore, we need to determine the values of the remaining $t-2$ thresholds.
The number of keys in a bin is binomially distributed.
To simplify analysis, we approximate the number of keys using a Poisson distribution with $\gamma k$ keys in expectation.
We consider the sorted fingerprints $0 < x_1 < x_2 < \dots$ of the keys.
The distance between any two consecutive keys ($x_{i+1}-x_{i}$) and between 0 and $x_1$ is exponentially distributed with an expected distance of $1/(\gamma k)$.
Hence, it is equivalent to obtain the fingerprints by sampling consecutive fingerprints using this exponential distribution until we pass 1.
However, for notational convenience we sample an infinite number of fingerprints in this way and allow them to exceed 1.
This does not change the distribution of the fingerprints smaller 1.

To minimize the number of empty slots we always choose the highest possible threshold that is still smaller than $x_{k+1}$, since we need to ensure that at most $k$ keys remain in the bin.
Hence, threshold $T_i$ is chosen if $T_i<x_{k+1}\leq T_{i+1}$ for $i\in [t]$ and we formally set $T_{t+1} := \infty$ to simplify notation.
A fingerprint $x_j$ results in an empty slot if it is bumped ($x_j>T_i$) \emph{and} it is one of the $k$ keys which we want to keep in the bin ($x_j<x_{k+1}$).
If we know $x_{k+1}$, then the $k$ smaller fingerprints are uniformly distributed in the range $(0, x_{k+1})$.
Hence, the expected number of empty slots is $k \frac{x_{k+1} - T_i}{x_{k+1}}$.
The fingerprint $x_{k+1}$ follows a gamma distribution with shape $k+1$ and rate $\gamma k$.
Let $\phi(x)$ be the density function of that gamma distribution.
We can calculate the expected number of empty slots of the bin as
\begin{align}
	\label{eq:bumptot}
	E := \mathbb{E}[\text{empty slots}] =& \int_{0}^{\infty} \mathbb{E}[\text{empty slots} ~|~ x_{k+1}=s] \phi(s)\mathrm{d}s\nonumber\\
	=& \sum_{i=1}^{t} \int_{T_i}^{T_{i+1}} \mathbb{E}[\text{empty slots} ~|~ x_{k+1}=s] \phi(s)\mathrm{d}s\nonumber\\
	=& \sum_{i=1}^{t} \int_{T_i}^{T_{i+1}} k \frac{s - T_i}{s} \phi(s)\mathrm{d}s
\end{align}
For $E$ to be minimal, its derivative with respect to $T_2, \dots, T_{t-1}$ must be zero.
We have:
\begin{align}
	0 \stackrel{!}{=} \frac{\mathrm{d}E}{\mathrm{d} T_i} &=
	\frac{\mathrm{d}}{\mathrm{d} T_i} \int_{T_{i-1}}^{T_{i}}k \frac{s - T_{i-1}}{s}\phi(s) \mathrm{d}s
	+
	\frac{\mathrm{d}}{\mathrm{d} T_i} \int_{T_{i}}^{T_{i+1}}k \frac{s - T_{i}}{s}\phi(s) \mathrm{d}s\nonumber\\
	&= k \left(
	\frac{T_i - T_{i-1}}{T_i}\phi(T_i)
	-
	\int_{T_{i}}^{T_{i+1}}\frac{1}{s}\phi(s) \mathrm{d}s
	\right)\nonumber\\
    &\iff \label{eq:optthresh}
	T_{i-1} =  T_i-\frac{T_i}{\phi(T_i)} \int_{T_{i}}^{T_{i+1}}\frac{1}{s}\phi(s) \mathrm{d}s
\end{align}
If we knew $T_t$ and $T_{t-1}$ we could inductively obtain the remaining thresholds.
However, we only know that $T_t=1$ and $T_1=0$.
Our implementation therefore fixes $T_t = 1$ and uses binary search of $T_{t-1}$ until $T_1$ is close to $0$.
This minimizes the number of empty slots under the necessary conditions $T_t=1$ and $T_1=0$.
For better intuition, we show \fvref{s:threshasymp} that the density of thresholds follows a gamma distribution for $t \to \infty$.
The query algorithm of our data structure is shown in \cref{alg:constructionQueryThreshold}.

\subsection{Tighter Packing With Retrieval}
Sometimes there is more than one key between the relevant thresholds.
If we take the smaller threshold, we keep empty slots that we later need to repair.
If we take the larger threshold, we would have more than $k$ keys in the bin.
In this case, we extend the data structure by querying a 1-bit retrieval data structure (see \cref{s:retrieval}) on all keys that have fingerprints between these two thresholds.
The bit tells us whether this key should be kept in the bin or if it should be bumped.
This costs us 1 bit per key for all keys between thresholds, even the ones we do \emph{not} bump.
However, overall, this reduces space because each bumped key leaving an empty slot costs us several bits of space later to repair.
The technique trades some query time for reduced space consumption.

\subsection{\consensus-Coded Thresholds}
\consensus \cite{lehmann2025consensus} is a technique for space-efficiently storing seeds for randomized data structures.
It basically uses a fixed number of bits to store the seed of each subtask.
This means that for some subtasks, there might be multiple successful seeds that can be represented with these bits, while for other tasks, there might be none.
The novel idea of \consensus is to concatenate the current seed with previous seeds.
This means that if we change one seed, we completely change the meaning of all following seeds.
That way, if one task has no successful seeds, we can backtrack to another successful seed in a previous task and get a completely new chance for success in the current task.

While the thresholds in threshold-based $k$-perfect hashing are not classical seeds, we can still apply the \consensus idea.
The similarity is that for some bins, we might have more than one threshold that gives a sufficiently small number of empty slots, while for other bins, there might be none.
We now use the thresholds of previous bins as a seed value when determining the fingerprints of the next bin.
Selecting a different threshold in a previous bin therefore gives us a new chance to reduce the number of empty slots.

A problem with this approach is that there might be some bins that are much larger than $\gamma k$ caused by uniformly hashing keys to bins.
Even if we try another hash function seed, the bins are biased towards having too many keys.
Therefore, finding a seed such that exactly $k$ keys are below one of the thresholds is unlikely.
However, we still have the option to bump too many keys and repair the empty slots later.
We need this option anyway because there might be bins that are not full enough.
Therefore, for each actual size of a bin, we determine how many keys we have to accept bumping if we want to have at least one valid threshold in expectation.
We then accept a seed if the number of empty slots is less than this limit.

\section{$k$-Perfect Hashing Through Bucket Placement}\label{s:bucketPlacement}
\begin{figure}[t]
    \centering
    \begin{subfigure}[b]{0.52\textwidth}
        \centering
        \includegraphics[scale=0.9]{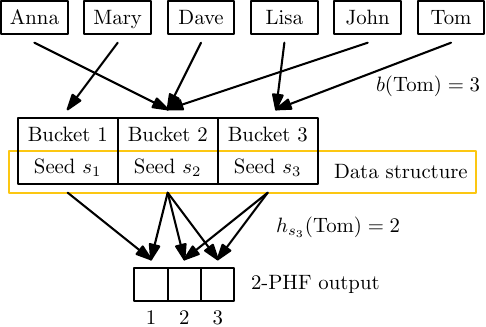}
        \subcaption{Illustration of the data structure using $k=2$.}
        \label{fig:bucketPlacementIllustration}
    \end{subfigure}
    \hfill
    \begin{subfigure}[b]{0.44\textwidth}
        \centering
        \input{fig/bucketfunction.tex}
        \subcaption{Bucket assignment functions for different values of $k$.}
        \label{fig:bucketfunction}
    \end{subfigure}
    \caption{Minimal $k$-perfect hashing through bucket placement.}
\end{figure}

Remember from \cref{s:relatedBucketPlacement} that perfect hashing through bucket placement \cite{pibiri2021pthash,hermann2024phobic,belazzougui2009hash} first hashes $n$ keys to $\frac{n}{\lambda}$ buckets.
For each bucket, it then searches for a hash function seed that can place the keys without collisions.
The \emph{bucket assignment function} $\beta: [0,1] \rightarrow [0,1]$ varies the expected size of each bucket such that placing each bucket has the same success probability.
\Cref{fig:bucketPlacementIllustration} gives an illustration of the idea and \cref{alg:constructionQueryBucket} describes the query algorithm.
The idea generalizes trivially to $k$-perfect hashing, where we now place a bucket if all output bins stay below $k$ collisions.
Because $k$-perfect hashing has different success probabilities than 1-perfect hashing, we need a different bucket assignment function $\beta_k$.
In this section we show how to choose $\beta_k$ such that each bucket has approximately the same success probability.

\begin{lstlisting}[label=alg:constructionQueryBucket,float=t,
caption={Query algorithm of $k$-perfect hashing trough bucket placement. Hash function $h_X$ uniformly maps into the set $X$.}]
Function query$(\textit{key} \in S)$
  $\textit{bucket}$ := $\lfloor (n/\lambda) \cdot \beta_k(h_{[0,1]}(\textit{key})) \rfloor$
  $\textit{seed}$ := &\textit{seeds}[$\textit{bucket}]$&
  return $h_{[\lceil n/k \rceil]}(\textit{key}, \textit{seed})$
\end{lstlisting}

We first need some general observations about bucket assignment functions.
For key $x$, the expected size of its bucket with normalized index $\beta_k(x)$ is $\lambda/\beta_k'(x)$.%
\footnote{In the close neighborhood of $x$ and for infinitesimal $\delta$, a $\delta$-fraction of the hash range (used by $\delta n$ keys in expectation) is shared by a $(\beta_k'(x) \cdot \delta)$-fraction of the $\frac{n}{\lambda}$ buckets. The quotient is $\delta n/(\beta_k'(x)\delta \frac{n}{\lambda}) = \lambda/\beta_k'(x)$ \cite{hermann2024phobic}.}
Let us assume that an $x$-fraction of keys is already mapped to bins.
Then let $p_k(x)$ be the probability that a random bin is not full at this point in the construction process.
When placing a bucket of size $s$, a seed is successful if all keys map to bins that are not full yet which has a probability of $p_k(x)^s$ (ignoring self collisions).
The optimal solution for $k=1$ satisfies that $p_k(x)^{\lambda/\beta_k'(x)}$ is some constant \cite{hermann2024phobic}.
Because $\lambda$ is a constant, the equation simplifies to $p_k(x)^{1/\beta_k'(x)} = C_k$.
Rearranging gives 
\begin{align}
	\label{eq:genbucket}
	\beta_k'(x) = \ln(p_k(x)) / \ln(C_k).
\end{align}
For $k=1$, a bin is full if there is a single key mapped to it.
We therefore have $p_1(x) = 1-x$ and $\beta_1'(x) = \ln(1-x)/ \ln(C_1)$.
The constant $C_k$ has to be chosen such that all keys are placed after the last bucket, i.e. $\beta_k(1) = 1$.
After integration and proper choice of $C_1$ we have $\beta_1(x) = x+(1-x)\ln(1-x)$, the optimal bucket assignment function for $k=1$ \cite{hermann2024phobic}.
The difficulty for $k>1$ is finding $p_k(x)$.
Once we have $p_k(x)$, we use \cref{eq:genbucket} to obtain the bucket assignment function.
Note that we assume without proof that using \cref{eq:genbucket} is still optimal for $k>1$.

\myparagraph{Model.}
To determine $p_k(x)$, we model the following process where we insert $bk$ keys into $b$ bins of capacity $k$.
We initialize a counter $c_i = 0$ for each bin $i \in [b]$.
The keys are handled sequentially as described in the following.
For each key we uniformly sample bins.
Sampling a bin $s \in [b]$ has two possible outcomes:
\begin{itemize}
	\item $c_s<k$ (\emph{Success}):
	We increment the counter $c_s$ by one and continue with the next key.
	\item $c_s\geq k$ (\emph{Failure}):
	In this case, we also increment the counter $c_s$ by one. Note that this does \emph{not} change anything in terms of success probabilities, since the bin is already full anyway. We continue with sampling a new bin for the same key.
\end{itemize}

\myparagraph{Determining $p_k(x)$ Numerically.}
The reason for incrementing the counter even if the bin is full is that the counter of each bin now follows a Poisson distribution (assuming that the number of bins $b$ is large).
We show how $p_k(x)$ can be expressed as the solution of an integro-differential equation \fvref{s:bucketdiff}.
However, solving this equation is numerically unstable.
In the following we give a numerically stable solution for $p_k(x)$.
We repeatedly choose values $\mu$.
We use each $\mu$ as the expected value of the Poisson distribution that describes the distribution of the counters.
For each $\mu$ we calculate the probability $p$ that a key is inserted successfully, as well as the fraction of inserted keys $x$.
Hence, for each $\mu$ we obtain $p$ and $x$ which we use to numerically determine $p_k(x)$.
We choose the values $\mu$ such that the distance of consecutive $x$ values is sufficiently small.
The success probability $p$ for a given $\mu$ is the probability that $c_s<k$, which we calculate using the cumulative distribution function of the Poisson distribution.
To determine the fraction of inserted keys $x$ for a given expected value $\mu$ we consider the counter of the bins as described in the model.
For a bin where the counter is higher than $k$, we charge $k$ successfully inserted keys, because all other incrementations are from failures.
We then divide by $k$, since we are interested in the relative number of inserted keys.
Let $X$ follow a Poisson distribution with mean $\mu$ then
$x = \frac{1}{k} (\mathbb{E}[X ~|~ X \leq k]+ k\mathbb{P}[X > k]) =\frac{1}{k} \mathbb{E}[X ~|~ X \leq k]+ \mathbb{P}[X > k]$.

\myparagraph{The Bucket Assignment Function.}
The final bucket assignment function can be obtained by numerically evaluating the integral of \cref{eq:genbucket} for $\beta_k(x)$.
The constant $C_k$ can be ignored by normalizing the integral as a last step such that the boundary condition $\beta_k(1)=1$ is met.
Refer to \cref{fig:bucketfunction} for an illustration of the bucket assignment functions with different values of $k$.
Larger $k$ make the function more aggressive, further increasing the size of the first buckets.
For example, for $k=100$, 82\% of the keys are hashed to 1\% of buckets.
Our implementation tabulates the bucket assignment function for specific $k$ and uses linear interpolation to maintain a fast query time.

\section{$k$-Perfect RecSplit}\label{s:recusiveSplitting}
RecSplit \cite{esposito2020recsplit} is a very space-efficient 1-perfect hash function.
We first explain RecSplit and then show how we adapt it to a $k$-perfect hash function.

\begin{figure}[t]
    \centering
    \includegraphics[scale=0.9]{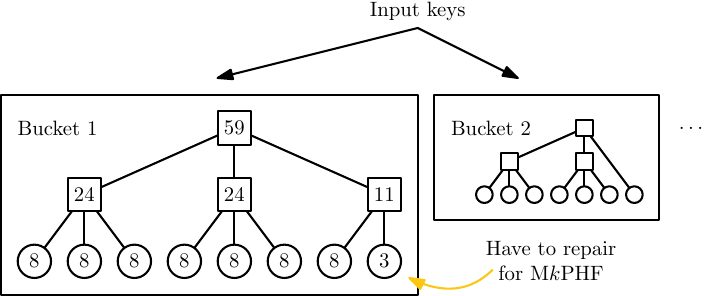}
    \caption{Illustration of RecSplit with $k=8$ and lowest-layer fanout $\ell=3$.}
    \label{fig:recsplitIllustration}
\end{figure}

\myparagraph{RecSplit.}
RecSplit \cite{esposito2020recsplit} first hashes each key to a bucket of expected size $b$, where $b$ is a tuning parameter with default value $b=2000$ in the implementation.
Within each bucket, RecSplit then uses brute-force to search for a hash function seed that divides the keys into two (or more) subsets.
The number of these subsets (\emph{fanout}) depends on the \emph{splitting strategy}.
This continues recursively until we are left with small input sets of $\leq \ell$ keys in the leaf nodes.
These subsets are small enough that it is feasible to search for a perfect hash function using brute-force.
RecSplit stores the hash function seeds determined at each node using Golomb-Rice coding (see \cref{s:golombRice}) in DFS order.
An Elias-Fano coded sequence (see \cref{s:eliasFano}) stores where the encoding of each bucket starts, as well as the total number of keys before it.
A query traverses the splitting tree of the corresponding bucket, evaluating the hash function in each node.
When descending into a subtree, it sums up the number of keys in sibling nodes left to it.
Therefore, the final hash function value is given by the number of keys before the bucket, the number of keys to the left in the tree, and the final brute-force hash function value in the leaf.
The RecSplit splitting strategy ensures that all subtrees except possibly the last are a complete tree where each leaf node receives exactly $\ell$ keys.
Therefore, only the very last leaf in any tree can have a size smaller than $\ell$.
We illustrate RecSplit in \cref{fig:recsplitIllustration}.
Through the use of the buckets, RecSplit achieves linear time construction and constant time queries.

\myparagraph{Main Idea.}
Our $k$-perfect adaptation of RecSplit is (almost) straight-forward.
We perform splittings until we are left with leaves of size $k$.
These then do not need to store seeds.
We now use the parameter $\ell$ to select the fanout of the layer above.
The main difficulty is that the last leaf in each bucket can have size $<k$.
This would make the result a non-minimal $k$PHF.
In \cref{s:nonKLeavesMerging}, we explain how we can merge the leaves of adjacent buckets in order to get output bins of size $k$.
As an alternative, in \cref{s:nonKLeavesNesting}, we explain \emph{nested} $k$-perfect hashing.
This can avoid the problem by ensuring that each bucket's size is a multiple of $k$.

\subsection{Merging Leaves of Size $< k$}\label{s:nonKLeavesMerging}
From the prefix sum of bucket sizes, we know the number of keys $x$ before each bucket.
We start placing the output of the bucket's tree at the offset $\lfloor x/k \rfloor$.
This means that we do not leave space for the previous bucket's leaves of size $<k$.
Only if the total number of keys in non-full leaves exceeds $k$, we get an additional output bin.
For this, we calculate a splitting hash function on the last leaf that splits away the keys needed to fill the leaf.
We store the splitting seed as an additional hash function seed in the Golomb-Rice coded seeds.

During queries, if we arrive in a leaf of size $<k$, there are two cases.
(1) If the remainder of the bucket position does not cross $k$, we scan forward the following buckets until one does.
The output bin is then the last bin of that bucket.
(2) If the current bucket is the one letting the remainder cross $k$, we evaluate the splitting hash function.
Then we either return the last bin of the current bucket, or we scan forward like in the first case.

Note that this means that we have to scan forward the bucket sizes when querying a leaf of size $<k$.
However, because the number of keys in a bucket is random and much larger than $k$, we can assume we get a uniform random remainder.
Therefore, in expectation, we only have to scan two buckets.

\subsection{Avoiding Leaves of Size $< k$ Through Nesting}\label{s:nonKLeavesNesting}
Another technique is to completely avoid leaves of size $< k$ in the first place.
We can use a minimal $b$-perfect hash function (where $k$ divides $b$) to partition the keys.
In general, we can nest different $k$-perfect hash functions to interpolate their trade-offs.
One function (maybe fast and less space-efficient) can map the input set to buckets of $k'$ keys.
Then, in each bucket, we can construct a $k$-perfect hash function.
We could use the idea to implement a $k$-perfect version of \consensus-RecSplit.
However, nesting different $k$PHFs would introduce another dimension of experiments and make this paper harder to follow.
We therefore do not go into detail about nesting or $k$-perfect \consensus-RecSplit.

\section{$k$-Perfect PaCHash}\label{s:kperfectPaCHash}

\begin{figure}[t]
    \centering
    \includegraphics[scale=0.9]{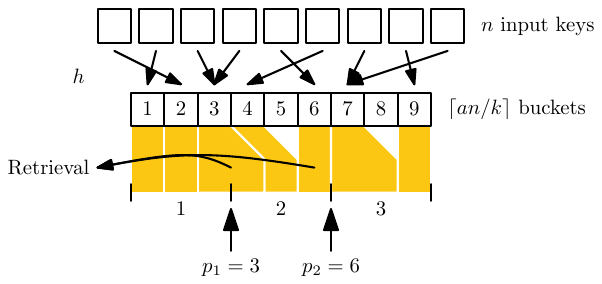}
    \caption{
        Illustration of PaCHash for $k$-perfect hashing with $a=k=3$.
        Queries for keys hashed to bucket 3 cannot directly determine whether their output bin is 1 or 2 by looking at the cut-points $p_i$.
        Therefore, they have to query a retrieval data structure.
        For the single key in bin 6, the placement would actually be unambiguous.
        However, from looking at $p_2$, we cannot see that all keys in bin 6 have to map to the left of the threshold.
        Therefore, we still need to query the retrieval data structure.
        For $p_2$, can store either 6 or 7.
        We select the bin with fewer keys to reduce the number of keys in the retrieval data structure.
    }
    \label{fig:pachashIllustration}
\end{figure}

As we explain in \cref{s:pachash}, PaCHash hashes each key to one of $\lceil an/k \rceil$ buckets.
The variable $a$ is a tuning parameter.
For each output bin $i$, it stores the first bucket $p_i$ overlapping it using Elias-Fano coding (see \cref{s:eliasFano}).
A predecessor query on that list then maps the bucket to the respective output bin.
\Cref{fig:pachashIllustration} illustrates the data structure.
During queries, if a searched bucket is contained in the sequence, it is not clear whether the key is supposed to be in the bin or in the bin before.
In fact, it might be possible that $d$ bins store the same bucket index.
We get a similar problem if a bucket starts exactly at a bin boundary.
Then the index cannot tell whether the bucket overlaps or not.
The external memory implementation then loads all $d+1$ candidate bins.
However, for $k$-perfect hashing, we need a different strategy.

\begin{lstlisting}[label=alg:queryPaCHash,float=t,
  caption={Query algorithm of $k$-perfect PaCHash. If the range returned by the internal memory data structure is ambiguous, we have to query the retrieval data structure. Hash function $h_X$ uniformly maps into the set $X$.}]
Function query$(\textit{key}\in S)$
    $b$ := $h_{[\lceil an/k \rceil]}(x)$
    find $i$ such that $p_{i-1} < b \leq p_i$ &\hfill&// &{\footnotesize predecessor query}&
    if $p_i = b$
      $i$ := $i-1$ &\hfill&// &{\footnotesize $b$ may start in previous block}&
    find first $j$ such that $p_j > b$  &\hfill&// &{\footnotesize predecessor query or scan}&
    if $i = (j-1)$
      return $i$
    else
      return $i + \textit{retrievalQuery}(\textit{key})$ &\hfill&// &{\footnotesize (j-1-i)-bit retrieval}&
\end{lstlisting}

\myparagraph{$k$-Perfect Adaptation.}
To decide for one of the $d+1$ bins using only the index data structure, we use a $\lceil\log_2(d+1)\rceil$ bit retrieval data structure (see \cref{s:retrieval}).
The stored value tells us which of the bins to return.
Since we rarely need more than one bit, we simply store the numbers bitwise in a single 1-bit retrieval data structure.
\Cref{alg:queryPaCHash} gives the pseudocode of the query algorithm.
PaCHash expects $1/a$ of the queries to load one external memory page too much \cite{kurpicz2023pachash}.
In the context of $k$-perfect hashing, this means that for $1/a$ of the queries, the output bin is ambiguous.
This means that the retrieval data structure essentially stores $n/a$ bits.
By selecting $a=k$, we get a total space consumption of $n/k(3 + \log(k))$, which already gets close to the lower bound for large $k$ (see \cref{s:spacebounds}).
An advantage of the technique is that its construction through (integer) sorting and a single scan is very simple.
Like the other approaches it takes linear time.
Queries take expected constant time due to the distribution of values in the Elias-Fano data structure \cite{kurpicz2023pachash}.
This adaptation is briefly mentioned in previous papers \cite{lehmann2024fast,lehmann2025consensus} but without an implementation.
The \consensus paper \cite{lehmann2025consensus} gives a simple analysis for the resulting space consumption.

\section{Evaluation}\label{s:eval}
In this section, we compare our approaches with CHD \cite{belazzougui2009hash}, as well as the ad-hoc implementation of threshold-based bumping \cite{lehmann2025consensus} from the literature.
While CHD is not \emph{minimal} $k$-perfect, we still include it in our comparison with a load factor of 97\% to give a feeling for the performance of our approaches.
With this paper, we build a foundation for future work on $k$-perfect hashing, contributing implementations for future papers to compare against.

We run our experiments on an Intel i7 11700 processor with 8 cores and a base clock speed of 2.5 GHz.
The machine runs Rocky Linux 9.5 with Linux 5.14.0.
We use the GNU C++ compiler version 11.2.0 with optimization flags \texttt{-O3 -march=native}.
Following the practice in 1-perfect hashing \cite{hermann2024phobic,lehmann2024fast, lehmann2025modern}, our input set consists of 100 million random strings of uniform random length $\in [10, 50]$.
Because all approaches hash the input keys anyway, the choice of the input set is not very important.
Our source code is public under the General Public License \cite{sourceCodekPHF}.

\begin{figure}[p]
    \centering
    \begin{tikzpicture}
        \ref*{legendPareto}
    \end{tikzpicture}

    \begin{subfigure}{\textwidth}
        \input{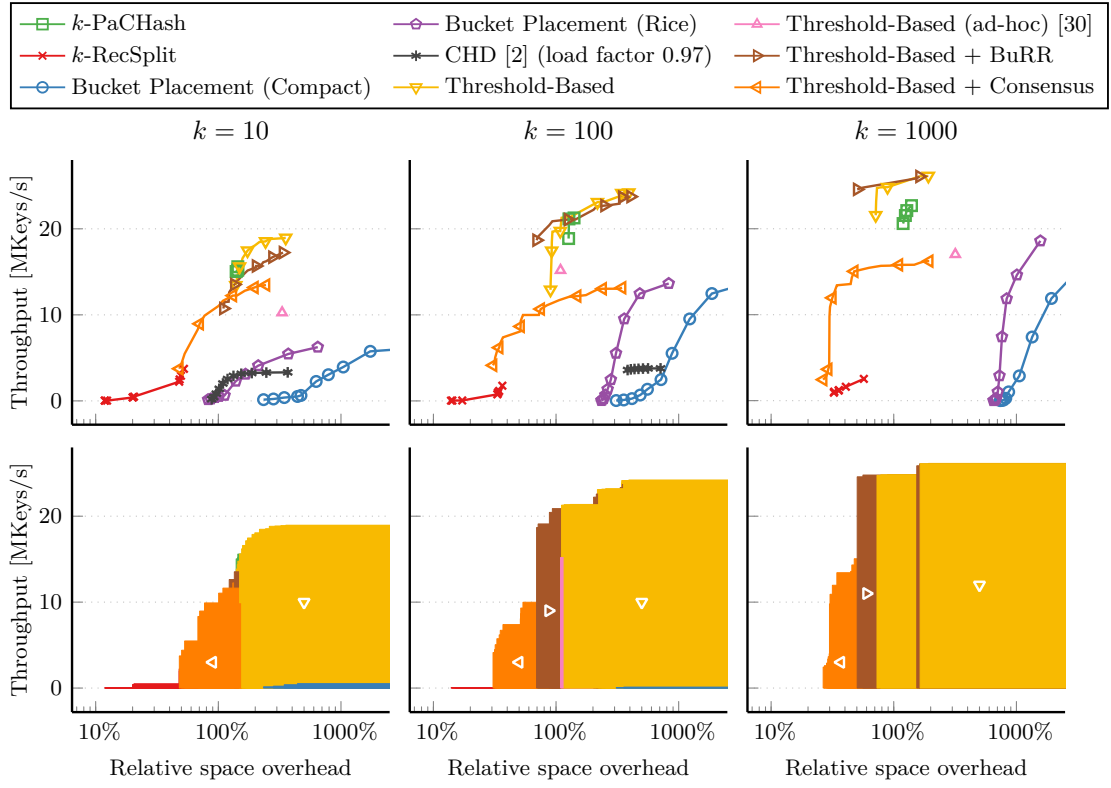}
        \subcaption{
            Space consumption versus construction time.
            The bottom row shows dominance maps indicating the approach with the fastest queries, given a specific trade-off between space and construction time.
        }
        \label{fig:paretoConstruction}
    \end{subfigure}
    \begin{subfigure}{\textwidth}
        \input{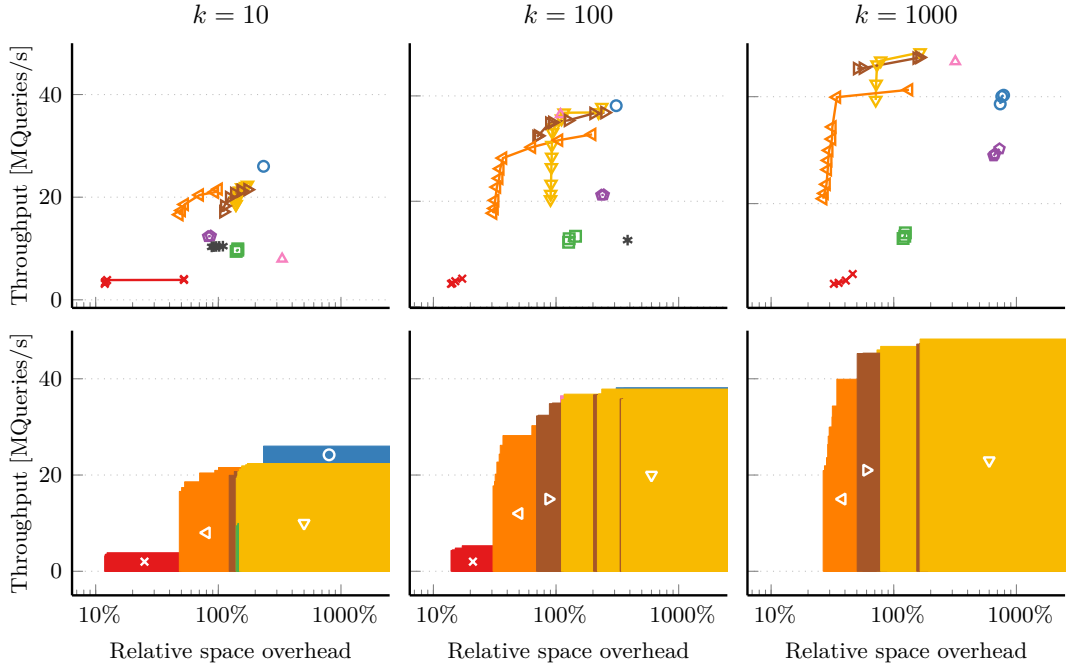}
        \subcaption{
            Space consumption versus query time.
            The bottom row shows dominance maps indicating the approach with the fastest construction, given a specific trade-off between space and query time.
        }
        \label{fig:paretoQueries}
    \end{subfigure}
    \caption{
        Comparison of minimal $k$-perfect hash functions.
        A relative space overhead of 100\% means that the data structure needs twice the lower bound.
        Refer to \cref{tab:bounds} for a table with lower bounds in bits per key.
    }
    \label{fig:pareto}
\end{figure}

Because each approach has configuration parameters, it can cover a range of trade-offs between space consumption, construction time, and query time.
To give an intuitive overview, we give visual plots showing different parameters of each approach.
\Cref{fig:paretoConstruction} gives the trade-off between construction time and space consumption for three different values of $k$.
\Cref{fig:paretoQueries} gives the trade-off between query time and space consumption.
We also give a \emph{dominance map} \cite{lehmann2024fast,dillinger2022burr, lehmann2025modern} of the trade-off.
For each point having a specific trade-off between construction time and space consumption, the dominance map in \cref{fig:paretoConstruction} shows the approach achieving the best query time.
As such, it can be seen as a ``front view'' of the 3-dimensional Pareto space.
The dominance map in \cref{fig:paretoQueries} shows the method with the fastest construction for a given trade-off between space consumption and query performance.
We show a selection of representative configurations from the Pareto front \fvref{s:selectedConfigurations}.
In the following, we discuss the different approaches in detail.

\myparagraph{Threshold-Based Bumping.}
Compared to the ad-hoc implementation \cite{lehmann2025consensus}, our implementation of threshold-based bumping offers up to 3 times smaller space overhead due to the optimized thresholds.
Only for $k=100$, the linearly spaced ad-hoc thresholds happen to be close to optimal, so we only achieve small space improvements.
For growing $k$, threshold-based bumping gets closer to the space lower bound, while the space overhead of most other approaches increases.
\Cref{fig:paretoConstruction} also shows that our implementation is up to 50\% faster to construct than the next best technique.
This is mainly because the ad-hoc implementation fully sorts the keys by their bin and fingerprint, while we only partition by the bin and then use quickselect \cite{hoare1961find} to find the $k$-th key.
By packing using retrieval, we can achieve smaller space consumption.
This has only a small influence on construction and query times.
Finally, using \consensus-coded thresholds, we achieve smaller space consumption but slower construction and queries.

\myparagraph{Hash and Displace.}
For the hash and displace technique, Rice coding (see \cref{s:golombRice}) achieves lower space consumption than compact coding at the same construction time.
This comes at the cost of slower queries, as can be seen in \cref{fig:paretoQueries}, due to the select queries.
Compared to CHD \cite{belazzougui2009hash}, an implementation from the literature, our construction is much faster.
Even though we use CHD with a load factor of just 97\%, our construction gets close to its space consumption.
The CHD implementation from the literature crashes for $k=1000$, while our implementation supports large $k$ as well.
Our measurements show that the hash and displace technique with compact coding is fast to query, especially for smaller $k$.
This is in contrast to the implementation from the literature, which has much slower queries.
In general, the hash and displace technique seems not competitive for $k$-perfect hashing, especially for large $k$.

\myparagraph{$k$-RecSplit.}
Our $k$-perfect adaptation of RecSplit achieves the smallest space consumption.
With just 10\% overhead over the lower bound, it has a large margin to the next smallest competitor.
This is not surprising, since the RecSplit paper \cite{esposito2020recsplit} shows that the recursive splitting idea converges towards the space lower bound.
However, its construction is significantly slower than the other approaches in \cref{fig:paretoConstruction}.
Because splittings become even more expensive for large $k$, we do not include the configuration with 10\% space overhead for $k=1000$.
The queries are slow compared to other approaches.

\myparagraph{$k$-PaCHash.}
Our $k$-perfect adaptation of PaCHash is very simple to describe and fast to construct through sorting followed by a linear scan.
However, its queries are slow due to the predecessor queries in the Elias-Fano coded sequence of bucket indexes.

\section{Conclusion}\label{s:conclusion}
To our knowledge, we present the first paper fully dedicated to presenting $k$-perfect hash functions.
The area of $k$-perfect hashing has been known for a long time, but has not received a lot of attention.
With the advent of very space-efficient 1-perfect hash functions, however, the area is becoming more and more relevant.
In this paper, we kickstart the development of $k$PHFs by introducing four new or significantly improved constructions.
Our constructions cover various trade-offs between space consumption, construction time, and query time.

With threshold-based bumping, we significantly improve a $k$-perfect hash function that was previously only described briefly as an ad-hoc solution in 1-perfect hashing papers \cite{lehmann2023bipartite,lehmann2025consensus}.
We achieve a significant reduction in bumped keys by combining it with retrieval, and we optimize the threshold function.
By combining the idea with \consensus coded seeds, we achieve a significant reduction in space consumption while retaining its fast queries.
We also extend the idea of perfect hashing through bucket placement by optimizing the bucket assignment function for $k$-perfect hashing.
As a very space-efficient alternative, we adapt RecSplit \cite{esposito2020recsplit} to construct $k$-perfect hash functions.
The main difficulty is to merge leaf nodes that have size~$< k$.
Finally, we adapt PaCHash \cite{kurpicz2023pachash} to a $k$-perfect hash function by combining it with a retrieval data structure.

Our extensive evaluation compares our new techniques with competitors from the literature.
Our constructions are simultaneously faster to construct, faster to query, and more space-efficient than previous approaches.
The most promising techniques are threshold-based bumping and $k$-RecSplit.
While $k$-RecSplit achieves the best space consumption, it is slow to construct and query.
Threshold-based bumping is very fast to construct and query while still giving competitive space consumption.

\myparagraph{Future Work.}
This paper revives the research area of $k$-perfect hashing.
In the future, we expect a wide range of new techniques, like there is in 1-perfect hashing.
The approaches presented in this paper give a starting point outlining the trade-off.
Especially our improvements to threshold-based bumping can now be used to improve the space efficiency of \consensus-RecSplit \cite{lehmann2025consensus}.
An open theoretical problem is the space lower bound for non-minimal $k$-perfect hashing.

\bibliography{paper}

\appendix
\crefalias{section}{appendix}

\section{Optimal Thresholds for \texorpdfstring{$t \to \infty$}{t to infinity}}
\label{s:threshasymp}
\begin{figure}[t]
	\centering
	\input{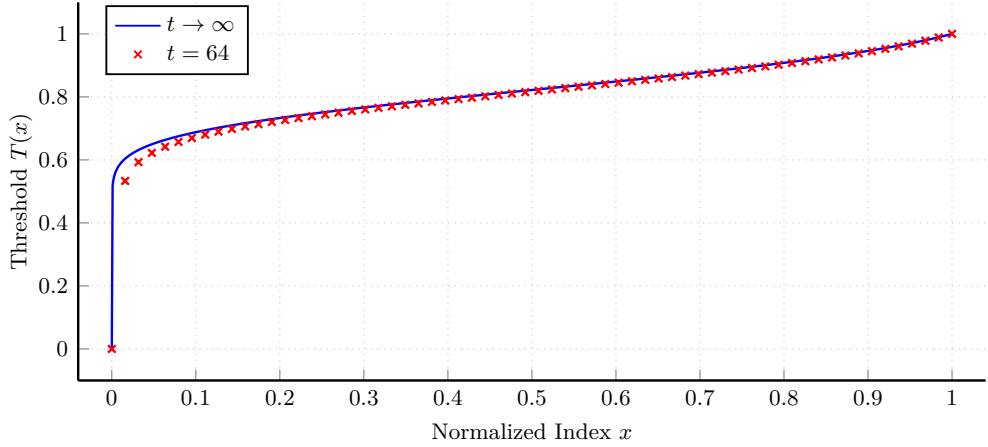}
	\caption{%
		Optimal thresholds for $k=100$ and $\gamma=1.2$. For an infinite number of thresholds we use the optimal threshold density as obtained in this section. For 64 thresholds we use the optimal solution obtained in \cref{ss:optthresh}.
    }
	\label{fig:thresh}
\end{figure}

In \cref{ss:optthresh} we determined optimal threshold values for $t$ thresholds using \cref{eq:optthresh}.
However, the formula is not intuitive.
In this section we consider how the thresholds are distributed for $t \to \infty$.
We define $T: [0,1] \rightarrow [0,1]$ such that $T_i = T(\frac{i}{t})$.
We basically express the threshold as a function of a normalized index.
Let $\phi_\alpha$ be a gamma distribution with shape $\alpha$ and rate $\gamma k$ where $\gamma$ is the overloading factor.
We rewrite \cref{eq:bumptot}, i.e. the expected number of empty slots as:
\begin{align*}
	E := \mathbb{E}[\text{empty slots}] &= \sum_{i=1}^{t} \int_{T_i}^{T_{i+1}} k \frac{s - T_i}{s} \phi_{k+1}(s)\mathrm{d}s \\
	&= l+k\sum_{i=1}^{t-1}\int_{T\left(\frac{i}{t}\right)}^{T\left(\frac{i+1}{t}\right)} \left(s - T\left(\frac{i}{t}\right)\right) \phi_k(s) \mathrm{d} s
\end{align*}
Where $l$ is the expected number of empty slots when the last threshold $T(1)=1$ is used.
The distance between consecutive thresholds approaches 0 for large $t$.
If this is not the case then the expected number of empty slots would not approach 0 in contrast to our optimal thresholds.
Note that $\phi_k(s)$ is continuous and the integral therefore simplifies to
\begin{align*}
\lim_{t\to\infty}	E =&l+k\sum_{i=1}^{t-1} \phi_k\left(T\left(\frac{i}{t}\right)\right) \int_{T\left(\frac{i}{t}\right)}^{T\left(\frac{i+1}{t}\right)} \left(s - T\left(\frac{i}{t}\right)\right) \mathrm{d} s \\
=& l+\frac{k}{2} \sum_{i=1}^{t-1} \phi_k\left(T\left(\frac{i}{t}\right)\right) \left( T\left(\frac{i+1}{t}\right) - T\left(\frac{i}{t}\right) \right)^2
\end{align*}
In the following we use the definition of the derivative
\begin{align*}
	T'(x) = \lim_{t\to\infty} t \left(T\left(x+\frac{1}{t}\right)-T(x)\right)
\end{align*}
to obtain
\begin{align*}
\lim_{t\to\infty}	E =& l+ \frac{k}{2t^2} \sum_{i=1}^{t-1} \phi_k\left(T\left(\frac{i}{t}\right)\right)  T'\left(\frac{i}{t}\right)^2
=d+ \frac{k}{2t} \int_{0}^{1} \phi_k(T(x)) T'(x)^2 dx
\end{align*}
We substitute $u=T(x)$ to obtain $du=T'(x)dx$, $x=T^{-1}(u)$ and
\begin{align*}
	\lim_{t\to\infty}	E =& l+ \frac{k}{2t} \int_{0}^{1}  \phi_k(u) T'(T^{-1}(u))  du
\end{align*}
We substitute $p(u) := \frac{1}{T'(T^{-1}(u))} = [T^{-1}]'(u)$ to obtain
\begin{align*}
	\lim_{t\to\infty}	E =  l+ \frac{k}{2t} \int_{0}^{1}\frac{\phi_k(u)}{p(u)} du
\end{align*}
We remark the similarity to the inverse transform sampling method: $p(u)$ describes the threshold density.
According to the Euler-Lagrange equation, $p(u)$ minimizes $\lim_{t\to\infty} E$ if
\begin{align*}
	0 \stackrel{!}{=} \frac{d}{du} \left( \frac{\partial}{\partial p} \frac{\phi_k(u)}{p(u)} \right)&= \frac{\phi_k'(u)p(u) - 2\phi_k(u)p'(u)}{p(u)^3} \\ &= \frac{\phi_{k-1}(u) \left((k (\gamma u-1)+1) p(u)+2 u p'(u)\right)}{p(u)^3}
\end{align*}
The differential equation is solved by choosing $p(u)$ proportional to a gamma distribution with shape $\frac{1}{2} (k + 1)$ and rate $\frac{1}{2} \gamma k$.
Finally, we normalize the density function in the range $[0,1]$.
Hence, the threshold density is just like the distribution of the key $x_{k+1}$, but with twice the variance.
Recall that $p(u) = [T^{-1}]'(u)$. We can use this to numerically determine the solution in terms of $T(x)$.
\cref{fig:thresh} shows our asymptotic result and the optimal solution for a finite number of thresholds as obtained in \cref{ss:optthresh}.
It can be seen that the solution for 64 thresholds is already close to the asymptotic result.
It is conceivable that sampling the threshold density function similar to the inverse transform sampling method is numerically more stable, faster and perhaps sufficiently close to the correct solution for a finite number of thresholds.
Future work might consider using the more simple sampling approach.

\section{Optimal Bucket Function as Integro-Differential Equation}
\label{s:bucketdiff}
Recall that in \cref{s:bucketPlacement} we determine an optimal bucket function and show that the number of keys in a bin follows a truncated Poisson distribution.
We assume the model as explained in \cref{s:bucketPlacement} and show an alternative approach to determine $p_k(x)$.
The expected value of the Poisson distribution is the average number of times a bin has been sampled.
The expected number of samples that need to be drawn until success for a key after an $x$ fraction of keys is inserted successfully is $1/p_k(x)$.
The expected value $\mu_k(x)$ of the Poisson distribution after an $x$ fraction of keys is inserted successfully is the sum of all samples until this point, distributed among the $\frac{n}{k}$ bins:
\begin{align}
	\label{eq:muk}
	\mu_k(x) = \lim_{n \to \infty} \frac{k}{n} \sum_{i=1}^{xn} \frac{1}{p_k(i/n)} = k \int_0^x \frac{1}{p_k(s)} \, ds
\end{align}
At the same time, the success probability $p_k(x)$ is the probability that the number of keys $X$ in a bin is less than $k$.
Since the number $X$ is Poisson distributed we have:
\begin{align}
	\label{eq:pkx}
	p_k(x) = \mathbb{P}[X \leq k - 1] = Q(k, \mu_k(x))
\end{align}
Where $Q$ is the regularized gamma function. Combining \cref{eq:pkx} and \cref{eq:muk} results in the integro-differential equation
\begin{align}
	\label{eq:bucket_integro}
	p_k(x) = Q \left(k, k \int_0^x \frac{1}{p_k(s)} \, ds \right)
\end{align}
Preliminary experiments show that solving \cref{eq:bucket_integro} using Euler's method is numerically unstable for large $k$.
Our implementation uses the numerically stable approach explained in \cref{s:bucketPlacement}.

\section{Additional Experimental Data}\label{s:selectedConfigurations}
\Cref{tab:parametersOverview} gives a table with a representative configuration of each approach.
Bucket placement achieves the fastest queries, but threshold-based bumping comes close.
For threshold-based bumping, packing with BuRR enables using a smaller number of thresholds while still reducing the space consumption.
The ad-hoc construction does not have configuration options.

\begin{table}[tp]
  \caption{%
      Selected configurations of all competitors.
  }
  \label{tab:parametersOverview}
  \centering

\newcommand{\rot}[2]{\multirow{#1}{*}{\rotatebox[origin=c]{90}{#2}}}
\addtolength\tabcolsep{-0.5pt}
\begin{centering}
\begin{tabular}[t]{lll rrr}
    \toprule
      & Approach & Configuration & Space    & Construction & Query \\
      &         &               & bits/key & ns/key       & ns/query \\ \midrule

      \rot{9}{$k=10$} &                           Bucket Placement (Compact) &              $\lambda$=12 & 1.917 & 584 &  47 \\
                      &                              Bucket Placement (Rice) &              $\lambda$=12 & 0.687 & 597 &  91 \\
                      &                                          $k$-PaCHash &                    $a$=10 & 0.733 &  67 & 103 \\
                      &                                      Threshold-Based & $\gamma$=2.0, $t$=$2^{5}$ & 0.716 &  77 &  56 \\
                      &                               Threshold-Based + BuRR & $\gamma$=2.0, $t$=$2^{4}$ & 0.630 &  96 &  61 \\
                      &                          Threshold-Based + Consensus & $\gamma$=2.0, $t$=$2^{4}$ & 0.450 & 233 &  57 \\
                      & Threshold-Based (ad-hoc) \cite{lehmann2025consensus} &                    $\bot$ & 1.293 &  97 & 137 \\
                      &    CHD \cite{belazzougui2009hash} (load factor 0.97) &             $\lambda$=$8$ & 0.627 & 495 &  96 \\
                      &                                         $k$-RecSplit &        $\ell$=2, $b$=2000 & 0.445 & 415 & 291 \\ \midrule
    \rot{8}{$k=1000$} &                           Bucket Placement (Compact) &             $\lambda$=250 & 0.073 & 361 &  25 \\
                      &                              Bucket Placement (Rice) &             $\lambda$=250 & 0.052 & 362 &  34 \\
                      &                                          $k$-PaCHash &                  $a$=1000 & 0.014 &  52 &  77 \\
                      &                                      Threshold-Based & $\gamma$=1.2, $t$=$2^{9}$ & 0.011 &  44 &  24 \\
                      &                               Threshold-Based + BuRR & $\gamma$=1.2, $t$=$2^{7}$ & 0.010 &  45 &  25 \\
                      &                          Threshold-Based + Consensus & $\gamma$=1.2, $t$=$2^{7}$ & 0.010 & 213 &  30 \\
                      & Threshold-Based (ad-hoc) \cite{lehmann2025consensus} &                    $\bot$ & 0.027 &  59 &  22 \\
                      &                                         $k$-RecSplit &        $\ell$=2, $b$=6000 & 0.010 & 417 & 172 \\
    \bottomrule
\end{tabular}
\end{centering}

\end{table}

\end{document}